\documentclass[pra,showpacs,twocolumn]{revtex4}
\usepackage{amsmath}
\usepackage{amssymb}
\usepackage{graphicx}
\usepackage{dcolumn}
\usepackage{bm}
\begin{document}
\title{A Manybody
Formalism for Fermions,
Enforcing the Pauli Principle on Paper}
\author{D.K.\ Watson \\
University of Oklahoma \\
Homer L.\ Dodge Department of Physics and Astronomy \\
Norman, OK 73019}
\date{\today}

\begin{abstract}
Confined quantum systems involving $N$ identical interacting
fermions are found in many areas of physics, including
condensed matter, atomic, nuclear and chemical physics. In a previous series 
of papers, a manybody
perturbation method that is applicable to both weakly
and strongly-interacting systems of bosons has been set forth
by the author and coworkers. A symmetry 
invariant perturbation theory was developed which uses group theory coupled
with the dimension of space as the perturbation parameter to obtain an
analytic correlated wave function through first order for 
a system under spherical confinement with a general two-body interaction.
 In the present paper, we extend
this formalism to large systems of fermions, circumventing the
numerical demands of applying the Pauli principle by enforcing the Pauli
principle on paper. The method does not
scale in complexity with $N$ and has minimal numerical cost.
We apply the method to a unitary 
Fermi gas and
compare to recent Monte Carlo values.
\end{abstract}
\maketitle

{\em Introduction.}-- Confined quantum systems of fermions are widespread
 across physics. 
They include, for example, atoms, atomic nuclei, neutron stars,
quantum dots, and cold Fermi gases.
These systems possess from a few tens to millions of
particles and span an enormous range of interparticle interaction strength
presenting a challenge for $N$-body methods
when mean-field approaches fail. In the last decade, ultracold Fermi gases have
emerged as a testing ground for manybody methods due to their precise 
controllability in experiments.  Using a magnetic field to tune the scattering 
length of atoms allows the exploration of the physics over many length scales,
including the ``unitary gas'' defined by an infinite scattering length. Without
a defining length scale other than the interparticle distance, the unitary
gas does not yield to conventional perturbation treatments, typically requiring
a full manybody treatment for an accurate description. 

Fermi systems in the unitary regime are currently of great interest. This
strongly interacting regime, stabilized by the Pauli exclusion principle, exists
on the cusp of the BCS-BEC crossover and exhibits universal thermodynamic
behavior which has been verified in the laboratory to within a few 
percent\cite{thermo}. 
Verifying this universal behavior theoretically requires
access to the partition function, i.e. the energy spectrum with
degeneracies. Required resources on
a classical computer for an {\it exact} solution of even the ground state 
of the $N$-body problem scale 
exponentially with $N$.
Traditional full configuration interaction can cope with at most some ten 
particles\cite{booth} and methods such as coupled-cluster\cite{CC} with a
computational time polynomial in $N$, $O(N^7)$, are extremely expensive.
Quantum Monte Carlo (QMC) methods for fermions suffer from the 
so-called ``sign''
problem\cite{mc,schmidt,gandolfi} resulting in an exponential growth in simulation times. 
Other methods that have been applied to 
strongly interacting
Fermi systems include the method of
correlated basis functions\cite{cbf,blumestechergreene}, density functional
theory\cite{singh}, diagrammatic approaches\cite{hu}, and the stochastic 
variational method \cite{varga,daily}. 
 To date, the determination of the energy spectrum of 
systems containing four or
more particles remains a challenge.\cite{daily} 

In this paper, we investigate an alternative approach to study large systems
of fermions.
Our symmetry-invariant perturbation theory, SPT, offers a perturbation approach
 for a  systematic study of correlation including the unitary regime.
The perturbation parameter depends on the inverse dimensionality of
space ($\delta = 1/D$), rather than the strength of the
interaction, so SPT is equally applicable to weakly or
strongly interacting systems. The method is essentially {\em analytic}
\cite{FGpaper}, with $N$, the number of particles,
entering as a parameter,   allowing
results for any $N$ to be obtained from a single
calculation\cite{energy}.
The lowest-order result
includes correlation and in theory, can be systematically 
improved by going to higher
order\cite{matrix_method}.
This method couples group theory with conventional dimensional perturbation
theory (DPT)\cite{copen92,loeser,chattrev} to take advantage of the high
degree of symmetry possible among identical particles in higher dimensions.
The terms in the perturbation
expansion are invariant under $N!$ symmetry operations, greatly reducing the 
number and complexity of the building blocks of these terms. This method
 currently includes
full manybody effects that are exact through first order.
Excited states are obtained
from the same analytic calculation\cite{energy}, differing only
in the number of quanta in the different normal modes. In principle,
the full energy spectrum is accessible.

In past work,  DPT  has been applied to fermion systems including single 
atoms\cite{loeser,rost}, quantum dots\cite{qdot}, and small 
molecules\cite{sung}.  The symmetry invariant 
approach, SPT, has been applied thus far only to 
bosons with spherical confinement, determining first-order energies, 
normal mode frequencies\cite{FGpaper,energy}, the lowest-order
SPT wave function\cite{paperI} and 
 density profile\cite{laingdensity}. We  
applied these results to a
BEC for which the density profile is a directly observable manifestation
of the quantized behavior. In a later series of papers, we
extended this work for bosons to first-order wave functions and density 
profiles\cite{JMPpaper,test}. We demonstrated that this method effectively 
rearranges the numerical work for this manybody problem into analytic building
blocks at each order giving the exact
result order by order in the perturbation series\cite{rearrangeprl,test}.
The complexity of the rearranged problem scales with the order of the
perturbation series,  not with the number of particles\cite{rearrangeprl}.

In this extension of SPT 
to fermions, we tackle the challenge of applying the Pauli principle, typically
an expensive numerical task. We describe how 
this is achieved ``on paper'' for any value of $N$,
 thus circumventing heavy numerical effort. We apply the method in this initial
 study to large systems of cold fermions in the unitary regime for which 
a number of very accurate calculations are available.  Results are 
obtained using analytic 
building blocks which have been calculated and stored previously and which
have been extensively checked using an independent solution of a model 
system of harmonically confined, harmonically interacting particles\cite{test}. Our
results compare well to very accurate Monte Carlo 
results\cite{blumestechergreene, chang} including some recent benchmark
calculations using the auxilliary field Monte Carlo method\cite{carlson}.

We begin the perturbation analysis by
defining dimensionally scaled quantities: $\bar{E} = \kappa(D) E,
 \mbox{and} \,  \bar{H} = \kappa(D)H$
%
where $\kappa(D)$ is a scale factor which
regularizes the large-dimension limit\cite{paperI}.
The scaled version of the Schrodinger equation becomes
\begin{equation} \label{eq:scale1}
\bar{H} \Phi =
\left(\frac{1}{\kappa(D)}\bar{T}+\bar{V}_{\mathtt{eff}}
\right)\Phi = \bar{E} \Phi,
\end{equation}
where barred quantities indicate variables in
scaled units ($\kappa(D) = D^2/(\hbar \bar{\omega}_{ho})$ for
this work. See Ref.~\cite{paperI}). The term $\bar{T}$ contains the derivative terms of the kinetic
energy and $\bar{V}_{\mathtt{eff}}$ includes centrifugal, two particle,
and confinement potentials\cite{paperI}.

We assume a totally symmetric, large-dimension configuration
at which the effective potential is a minimum.  The $N$
particles are arranged on a hypersphere, each particle with a radius,
$\bar{r}_{\infty}$, from the center of the confining potential.
Furthermore, the angle cosines between each pair of particles take
on the same value, $\overline{\gamma}_{\infty}$, i.e.\
\begin{equation} \label{eq:Lewis}
 \renewcommand{\arraystretch}{1.5}
\begin{array}{rcl}
{\displaystyle \lim_{D \rightarrow \infty} \bar{r}_{i} } & = &
{\displaystyle \bar{r}_{\infty} \;\; (1 \le i \le N)\,, } \\
{\displaystyle \lim_{D \rightarrow \infty} \gamma_{ij} } & = &
{\displaystyle \overline{\gamma}_{\infty} \;\; (1 \le i < j \le
N)\,. }
\end{array}
 \renewcommand{\arraystretch}{1}
\end{equation}

\noindent (This symmetric high-dimensional structure is not unlike the
localized structure found in a hyperspherical treatment of the
confined two-component Fermi gas in the $N \rightarrow
\infty$ limit\cite{seth}.)
In scaled units the  
$\delta \rightarrow 0$ ($D \rightarrow \infty$) 
approximation for the energy is
simply the effective potential minimum, i.e.
$\bar{E}_{\infty}=\bar{V}_{\mathtt{eff}}(\bar{r}_{\infty},\overline{\gamma}_{\infty};
\hspace{1ex} \delta=0)\,$.

In this $\delta \to 0$ approximation, the centrifugal-like
term that appears in $\bar{V}_{\mathtt{eff}}$\,, which is nonzero even for the
ground state, is a zero-point energy contribution satisfying the
minimum uncertainty principle\cite{chat}. The value of
$\overline{\gamma}_{\infty}$\,,
 which is zero
in the {\em mean-field}
approximation for the $L=0$ angular momentum states considered here, is,
in fact, {\em not} zero, an
indication that beyond-mean-field effects are included in the
$\delta \to 0$ limit.

 This highly-symmetric, $\delta \to 0$ structure imparts a
point group structure to the system which is isomorphic to the
symmetric group of $N$ identical objects\cite{hamermesh}, $S_N$, allowing
a largely analytic solution.
The $\delta \to 0$ approximation may be systematically
improved by using it as the starting point for a perturbation
expansion\cite{matrix_method}. The $S_N$
symmetry greatly simplifies this task since the interaction terms
individually have to transform as a scalar under the $S_N$ point
group.

The perturbation series has the form:
\begin{equation} \renewcommand{\arraystretch}{2} \begin{array}{r@{}l@{}c}
{\displaystyle \bar{E} = \bar{E}_{\infty} + \delta} &
{\displaystyle \, \sum_{j=0}^\infty
\left(\delta^{\frac{1}{2}}\right)^j \,
\bar{E}_j } & \\
{\displaystyle \Phi = } & {\displaystyle \,
\sum_{j=0}^\infty \left(\delta^{\frac{1}{2}}\right)^j \, \Phi_j }& \,.
\end{array}
\renewcommand{\arraystretch}{1}
\end{equation}
In practice $\bar{E}_j=0$ $\forall$ $j$ odd. The $j=0$ terms
are obtained from a harmonic equation, and referred to
as the energy and wave function at harmonic
order.
To obtain this harmonic correction for small values of $\delta$\,, 
we expand about the minimum of the
$\delta \to 0$ effective potential.

The harmonic-order Hamiltonian
is solved using the FG matrix method\cite{dcw} 
to obtain the normal-mode frequencies, $\bar{\omega}_{\mu}$.
The number of roots, $\lambda_{\mu}$, 
($\lambda_{\mu}={\bar{\omega}_{\mu}}^2$)\,, of the secular equation,
 $N(N+1)/2$ roots, is potentially huge;
however, due to the $S_N$ symmetry of the problem 
there is a reduction to five distinct roots,

The $\bm{FG}$ matrix is
invariant under $S_N$\,, so it does not connect subspaces
belonging to different irreducible representations (irreps.) of
$S_N$\cite{WDC}. Thus
the normal coordinates must transform under irreps. of $S_N$\,.
 The normal coordinates are
linear combinations of the elements of the internal 
displacement vectors which 
transform under reducible representations of $S_N$\,.
One can show that these reduce to two $1$-dimensional $[N]$
irreps. denoted by ${\bf 0^+, 0^-}$, two
$(N-1)$-dimensional $[N-1,1]$ irreps. denoted by ${\bf 1^+,1^-}$, 
and one angular
$N(N-3)/2$-dimensional $[N-2,2]$ irrep. 
denoted by $\bf 2$\cite{FGpaper}.

The energy through harmonic order in $\delta$ is
\cite{FGpaper}
\begin{equation}
\overline{E} = \overline{E}_{\infty} + \delta \Biggl[
\sum_{\renewcommand{\arraystretch}{0}
\begin{array}[t]{r@{}l@{}c@{}l@{}l} \scriptstyle \mu = \{
  & \scriptstyle \bm{0}^\pm,\hspace{0.5ex}
  & \scriptstyle \bm{1}^\pm & , & \\
  &  \,\scriptstyle \bm{2}   \scriptstyle  \}
            \end{array}
            \renewcommand{\arraystretch}{1} }
(n_{\mu}+\frac{1}{2} d_{\mu})
\bar{\omega}_{\mu} \, + \, v_o \Biggr] \,, \label{eq:E1}
\end{equation}
where $n_{\mu}$ is the total number of quanta in the normal mode
with the frequency $\bar{\omega}_{\mu}$;
 $\mu$ is a label which
runs over ${\bf 0}^-$\,, ${\bf
0}^+$\,, ${\bf 1}^-$\,, ${\bf 1}^+$\,, and ${\bf 2}$\,, regardless
of the number of particles in the system (see Refs.~\cite{FGpaper}and Ref.[15]
in \cite{paperI}), and $v_o$ is a constant.
The
multiplicities of the five roots are:
$d_{{\bf 0}^+} = 1, \hspace{1ex} d_{{\bf 0}^-} = 1,\;
d_{{\bf 1}^+} = N-1,\;  d_{{\bf 1}^-} = N-1,\;
d_{{\bf 2}} = N(N-3)/2$.

A character analysis of the normal modes reveals that
the ${\bf 2}$ normal modes are phonon, i.e. compressional modes;
the ${\bf 1^\pm}$ modes show single-particle character, and the
${\bf 0^\pm}$  modes 
describe center-of-mass and breathing motions.


{\em Enforcing the Pauli Principle.}-- To generalize SPT from
quantum systems of bosons to quantum systems of fermions, we must enforce the
Pauli principle, thus requiring the $N$-body wave function to be totally
antisymmetric.  This is enforced by placing certain restrictions on the 
occupancies of
the normal modes, i.e. on the values of the normal mode quantum numbers,
$n_{\mu}$, $\mu = {\bf 0}^\pm,{\bf 1}^\pm,{\bf 2}$ in  
Eq.~(\ref{eq:E1})\cite{loeser}. 
 The possible assignments can be found by
relating the normal mode states 
$|n_{{\bf 0}^+},n_{{\bf 0}^-},n_{{\bf 1}^+},n_{{\bf 1}^-},n_{\bf 2}>$ to the
states of the confining potential which is a spherically symmetric
three dimensional harmonic oscillator 
$(V_{\mathtt{conf}}(r_i)=\frac{1}{2}m\omega_{ho}^2{r_i}^2)$
 for which the restrictions
imposed by antisymmetry are known.  These two series of states can be related 
in the double limit  $D\to\infty$, $\omega_{ho}\to\infty$ where both
representations are valid.

For large $D$, the normal mode description given by  Eq.~(\ref{eq:E1})
is exact. Applying the large $\omega_{ho}$ limit results in:

\begin{equation}
E = N\frac{D}{2} \hbar {\omega_{ho}} + (2 n_{{\bf 0}^+} + 2 n_{{\bf 0}^-} 
+ 2 n_{{\bf 1}^+} + 2 n_{{\bf 1}^-} + 2 n_{\bf 2}) \hbar {\omega_{ho}}
\end{equation}

Now consider   $\omega_{ho}
\rightarrow \infty$ first and then $D \rightarrow \infty$.
The harmonic oscillator levels are 
exact:

\begin{equation}
\label{Elargeomega}
\renewcommand{\arraystretch}{1.5}
\begin{array}[b]{r@{\hspace{0.4ex}}c@{\hspace{0.2ex}}l} 
{\displaystyle E} &=&  
{\displaystyle \sum_{i=1}^N
\left[(2\nu_i + l_i) +\frac{D}{2}\right] \hbar\omega_{ho}}
\\ &&
= {\displaystyle N\frac{D}{2} \hbar\omega_{ho} + 
\sum_{i=1}^N (2\nu_i + l_i) \hbar \omega_{ho}}
\end{array} \renewcommand{\arraystretch}{1}
\end{equation}

\noindent where  $\nu_i$
is a radial quantum number and $l_i$ the orbital angular momentum
quantum number. 
Equating these two expressions which are equal in the double limit, the
quantum numbers in the two representations can now be related to show the
restrictions on normal mode states imposed by antisymmetry.  Because of the
clean separation of radial and angular motions, two conditions result:

\begin{equation} \renewcommand{\arraystretch}{1} \begin{array}{r@{}l@{}c}
{\displaystyle 2 n_{{\bf 0}^-} + 2 n_{{\bf 1}^-} } =  &
{\displaystyle \, \sum_{i=1}^N 2 \nu_i \,
 } & \\
{\displaystyle  2 n_{{\bf 0}^+} + 2 n_{{\bf 1}^+} + 2 n_{\bf 2} = } & 
{\displaystyle \,
\sum_{i=1}^N  l_i }& \,.
\end{array}
\renewcommand{\arraystretch}{1}
\end{equation}

\noindent These equations determine a set of possible normal mode states
$|n_{{\bf 0}^+},n_{{\bf 0}^-},n_{{\bf 1}^+},n_{{\bf 1}^-},n_{\bf 2}>$
 from the known set of permissible
$L=0$ harmonic oscillator configurations.

{\em Application: The Unitary Gas.}-- The Schr\"odinger equation for an
$N$-body system of fermions, $N=N_1+N_2$ with $N_1$ spin up and
$N_2$ spin down fermions, confined by a spherically symmetric
potential is
\begin{equation}
\label{generalH} H \Psi = \left[ \sum\limits_{i=1}^{N} h_{i} +
\sum_{i=1}^{N_1}\sum\limits_{j=1}^{N_2} g_{ij} \right] \Psi = E
\Psi \,.
\end{equation}
\noindent where $h_i$ and $g_{ij}$
are the single-particle Hamiltonian and the two-body interaction
potential, respectively.
We assume a $T=0K$
condensate  with $N_1 =N_2$ confined by an isotropic, harmonic trap with
frequency $\omega_{ho}$.

 To study the unitary regime, we replace the actual atom-atom
potential by an attractive square well potential of radius $R$:
\begin{equation}
V_{\mathtt{int}}(r_{ij})=
\left\{ \begin{array}{ll} -V_o\,, & r_{ij} < R \\
\,\,0,& r_{ij} \ge R\,.
\end{array}\right. \,,
\end{equation}
For fixed range $R$, the potential depth $V_0$ is adjusted so the 
s-wave scattering length, $a_s$ is infinite. The range is selected so
$R<<a_{ho}$ ( $a_{ho}=\sqrt{\hbar/(m\omega_{ho})}$)  
and can be systematically reduced to extrapolate 
to zero-range interaction.
We dimensionally continue the square well potential so that it is
differentiable away from $D=3$\,, allowing us to perform the
dimensional perturbation analysis\cite{FGpaper,energy}).
 Thus, we take the
interaction to be
\begin{equation} \label{eq:vint2}
V_{\mathtt{int}}(r_{ij})= V_0(\delta) \left[1-\tanh \left[
\vphantom{\left(r_{ij}-\alpha-\frac{3}{D}(a-\alpha)\right)}
\frac{1}{1-3\delta} \left(r_{ij}-3\delta R \right) \right] \right] \,,
\end{equation}

\noindent where $ V_o(\delta)=\frac{1}{1-3b\delta}$. The potential depth $V_o$ is adjusted
by adjusting the value of $b$ so the scattering length
is infinite when $\delta = \frac{1}{3}$.
 This
interaction becomes a square well of radius $R$ in the physical
$D=3$\, limit.   The functional
form of the potential at $D\neq 3$ is not unique. Other 
forms could be chosen with equal success as long as the form is
differentiable and reduces to a square well potential at $D=3$\,.
We simply choose a form that allows a gradual softening of the
square well.

%
\begin{figure}
\includegraphics[scale=.6]{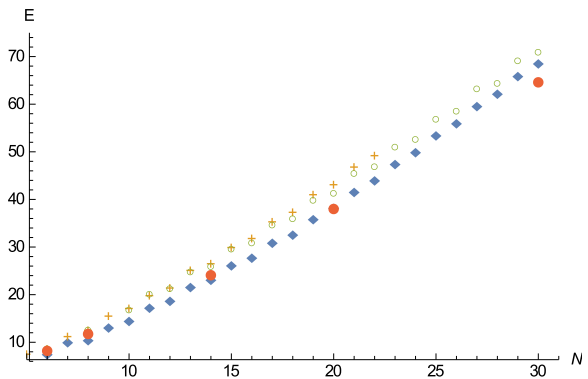}
\renewcommand{\baselinestretch}{.9}
\caption{Ground state energies of the harmonically trapped unitary Fermi gas
(units $\hbar{\omega}_{ho} = 1$). Our first-order perturbation results 
(filled diamonds) are compared to GFMC (+'s)\cite{chang}, fixed-node 
DMC (open circles) from Ref. \cite{blumestechergreene}, and 
AFMC results (filled circles, $N$=6, 8, 14, 20, 30 only)\cite{carlson}.
} \label{fig:one}
\end{figure}

\begin{figure}
\includegraphics[scale=.6]{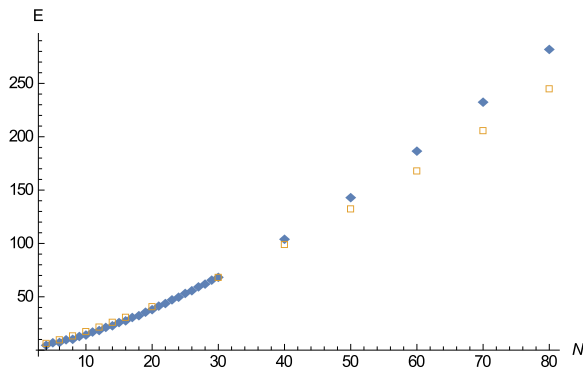}
\renewcommand{\baselinestretch}{.9}
\caption{Ground state energies of the harmonically trapped unitary Fermi gas
(units $\hbar{\omega}_{ho} = 1$). Our first-order perturbation results 
(filled diamonds) are compared to DMC results Ref. \cite{carlson}
(open squares)
} \label{fig:two}
\end{figure}
In Fig.~\ref{fig:one} we plot the ground state energies  
from $N=6$ to $N=30$ and compare to Green's function Monte Carlo (GFMC)
 energies\cite{chang},  
to fixed-node diffusion Monte Carlo (DMC)
energies\cite{blumestechergreene} which provide accurate
upper bounds to the ground state energy, and to
recent benchmark auxiliary field Monte Carlo 
calculations (AFMC)\cite{carlson} 
which are exact, but subject to finite lattice size errors for which
 corrections have been made. The AFMC results are currently the most accurate
results available.
 Our energies which include full manybody effects through
first order compare well to these AFMC results and except
for $N = 8$ they are slightly closer to the AFMC results
than the GFMC and DMC results
shown in Figure 1. Our numbers, as well as the GFMC and DMC numbers from 
Refs.~\cite{chang, blumestechergreene}, show a distinct odd/even oscillation, 
but no 
obvious shell effects. Obtained from analytic building blocks
that have been calculated and stored previously, our numerical
requirements take a few seconds on
a work station. 

No parameters
are used to obtain these results other than those used to produce a square
well potential with infinite scattering length. For the GFMC and DMC studies
 shown
in Figure 1, no results exist for values of $N$ higher than shown. 
Accurate results 
for higher
$N$ are, of course, increasingly difficult to achieve. In Figure 2 we compare
our results at first order
above $N = 30$ with available DMC results from Ref.~\cite{carlson}.
Our first-order results show the expected increase in error as N increases 
suggesting that for larger N, higher order terms may be necessary.

{\em Conclusions} -- In this paper we have extended the symmetry-invariant 
perturbation method
from bosons to fermions, applying the Pauli principle ``on paper'' to avoid
heavy numerical expense. The method has been tested in the unitary regime,
which is of particular interest for manybody methods since its infinite
scattering length and the lack of a natural scale typically require intensive
numerical simulation for an accurate description.  Our analytic results through
first order yield energies
that are comparable in accuracy with 
recent Monte Carlo results. As $N$ increases, our error  
increases suggesting the need for higher order terms.
 It may also be possible to rearrange the perturbation 
series to minimize the importance of higher order terms.
 
The theory applied in this paper is applicable to 
$L = 0$ states of spherically confined systems 
with general attractive or repulsive interparticle interactions
and is also applicable to both
weakly and strongly correlated
systems.
The fact that
$\overline{\gamma}_\infty$ is not zero is an indication that
beyond-mean-field effects are included in this result even in the
$D\rightarrow\infty$ limit.
This theory is readily
generalizable to systems with a cylindrical confining potential.

Higher-order calculations may be required
for larger $N$. A detailed program for calculating
higher-order DPT corrections to $N$-body systems
has been laid out\cite{matrix_method} 
and applied to high order for
small-$N$ systems\cite{highorder}. For
large-$N$ systems the $S_N$ point-group symmetry 
 greatly simplifies the calculation of these higher-order
terms.


We thank Doerte Blume and Stefano Gandolfi for discussions and
Monte Carlo results.

\end{document}